\documentclass{elsart}

\usepackage{natbib}

\usepackage{amssymb,psfig}

\begin{document}

\begin{frontmatter}



\title{Potential Science for the OASIS Integral Field Spectrograph with Laser Guide Star Adaptive Optics}


\author[Dur]{Simon L. Morris}
\ead{simon.morris@durham.ac.uk}
\author[Dur]{Joris Gerssen}
\author[Dur]{Mark Swinbank}
\author[Dur]{Richard Wilman}

\address[Dur]{Department of Physics, University of Durham, Rochester Building,
Science Laboratories, South Road, Durham DH1 3LE, UK}

\begin{abstract}

I review the science case for the Laser Guide Star system being
built for the William Herschel Telescope (WHT) on La Palma. When
used in combination with the NAOMI Adaptive Optics system and the
OASIS visible-wavelength Integral Field Spectrograph, I demonstrate
that there are substantial, exciting areas of astrophysical research
in which the WHT can contribute.

\end{abstract}


\end{frontmatter}

\section{Introduction}
\label{sec-intro}

Many people have been involved in generating the science case for
the addition of a Rayleigh Laser Guide Star (LGS)
\citep{2003SPIE.4839..360R, 2003SPIE.4839..516C,
2004INGN....8...13R} to the OASIS IFU spectrograph
\citep{2003INGN....7...21B, 2004INGN....8....3M,
2004SPIE.5492..822M} combined with the NAOMI Adaptive Optics (AO)
system \citep{2003SPIE.4839..647M, 2004SPIE.5490..280J,
2004SPIE.5490...79B} on the William Herschel Telescope on La Palma.
The co-authors listed on some versions of the LGS proposal are: Rene
Rutten, Roland Bacon, Richard Bingham, Paul Clark, Roger Davies, Ron
Humphreys, Tom Gregory, Johan Knapen, Gil Moretto, Tim Morris, Simon
Morris, Richard Myers, Gordon Talbot, Richard Wilson, Tim de Zeeuw.
As a member of this group, I (SLM) was invited to summarize this
science case. I have chosen to only cover part of the science case,
and also (as is traditional for conference reviewers the world over)
to focus on science areas currently being pursued by myself or
colleagues at Durham University. As a result, this is by no means an
exhaustive list, but it is one that I am better qualified to write
on, and it does contain some updates from the original proposal.

The topics discussed below will all be ones that can be attacked
with the instrument combination in the title (i.e. in the optical,
with relatively low Strehl), and using the WHT collecting area,
although obviously many of them are also topics that larger aperture
telescopes are also addressing. I have also tried to focus on cases
where observations with the WHT are not only useful, but also where
the OASIS+NAOMI+LGS combination there has some sort of an advantage
or at least only a small number of competitors.

\section{Science Changes produced by Installation of a Laser Guide Star}
\label{sec-changes}

This subject was covered by several people attending the workshop,
but in order to make this summary relatively self-contained, I will
repeat some of it here.

Far and away the most significant change in the science potential
for OASIS+NAOMI from the addition of the LGS is the gain in sky
coverage. Some graphical demonstrations of the magnitude of this
change were developed by Remko Stuik (Leiden), and shown in several
of the presentations. Apart from deep in the galactic plane, the
chance of finding a bright enough guide star for NAOMI without a LGS
is never higher that 10\%. In contrast, with an LGS, the probability
of finding a suitable natural tip-tilt star (which is still needed)
is above 80\% for 2/3 of the sky, and above 60\% for near 9/10 of
the sky. As will be detailed below, this makes a qualitative
difference to the types of science that can be done. Targets which
are suited to 4m class high spatial resolution IFU spectroscopy are
rare, and multiplying that rarity by the small sky coverage of an
NGS AO system reduces the number of interesting science targets to
near zero.

Some secondary effects are of mixed value. The LGS point spread
function (PSF) can either be better or worse than the NGS PSF
depending on the NGS brightness and location. For many science
targets this means there will be an improvement in the delivered
image quality, even if a usable NGS is in the field. For a few
objects this will not not be the case. In all cases the PSF will be
very far from the diffraction limit due to operation in the optical
with a (relatively) modest number of correcting elements. As was
emphasized during the meeting, and in some of the examples below,
this sort of 'improved seeing' can be extremely valuable. The
predictable brightness and location of the LGS relative to the
science target will make the science target PSF more predictable and
reduce the challenges of estimating the delivered image quality.
Finally, the operational impact of keeping an LGS up and running may
or may not impact on the efficiency of observing. Overheads could be
increased by the need to lock on to the laser, but, conversely,  its
predictable and repeatable location and brightness could produce
efficiency savings.

\section{An Example of Galactic Science Enabled by a LGS}
\label{sec-galactic}

In the proposal, galactic science targets ranging from sub-stellar
companions, through mass loss in stars, to exotic binary systems in
crowded fields were considered. Here I will focus on one
particularly intriguing science case - namely the search for
intermediate mass black holes.

\subsection{Intermediate Mass Black Holes in Globular Clusters}
\label{subsec-imbh}

Although the evidence for stellar mass black holes (M$\le10
M_{\odot}$) and for supermassive black holes (M$\ge10^6 M_{\odot}$)
is now viewed by most astronomers as iron-clad, there is a `missing
link' mass range in between for which there is little observational
support. One very promising hunting ground for black holes in this
mass range is the core of a globular cluster. This is for two
reasons. First, that many formation scenarios for such black hole
masses place them at the bottom of potential wells in crowded
stellar environments, and second, that the crowded stellar
environment provides one with a surplus of test masses which can map
out the potential well and be used to test for point gravitational
sources.

Tragically, this same surplus of test particles makes crowding a
real problem, and deblending the light from individual stars in
order to have a clean velocity measurement requires high spatial
resolution and a good knowledge of the PSF.

Valiant attempts to conclusively demonstrate the presence of an
intermediate black hole were made by \citet{2002AJ....124.3255V,
2002AJ....124.3270G, 2003AJ....125..376G} and
\citet{2004AN....325...84G}. As an example, using the Hubble Space
Telescope (HST) STIS spectrograph, radial velocities for 64 stars in
the core of the globular cluster M15 were measured and combined with
velocities measured from the ground for 1800 other stars.
Unfortunately, although it was possible to show that a model with a
1700 $M_{\odot}$ black hole gave a better fit to the data than one
without one, the `no black hole' model was still statistically
consistent with the data.

An AO fed IFU able to observe the cores of several nearby globular
clusters may well be able to improve the statistics to the point at
which the black hole solution is shown to be inescapable. However,
detailed modelling is still needed to show that (a) the OASIS
spectral resolution is sufficient, and (b) that the PSF improvements
from NAOMI+LGS in the optical are sufficient for this approach to
generate new science.

\section{Examples of Nearby-Galaxy Science Enabled by a LGS}
\label{sec-neargal}

The workshop poster carried a dramatic image illustrating the
additional science to be gained by increasing the spatial resolution
of velocity measurements in the cores of nearby galaxies
(http://www.ast.cam.ac.uk/ING/conferences/aoworkshop/aoworkshop.php).
This image was literally the `poster-child' for the workshop and was
very persuasive. I do not propose to go into this science case in
any detail, but do include a few references here for completeness.

\subsection{Detailed Dynamics of Nearby Spheroid Dominated Galaxies}
\label{subsec-sauron}

The SAURON IFU spectrograph had an extremely well defined and
focussed science goal. As I will outline later, this has not stopped
it from being used for a number of other science projects, but it
has certainly amply achieved its builders' goals as described in
\citet{2001MNRAS.326...23B, 2002MNRAS.329..513D} and
\citet{2004MNRAS.352..721E}. The core goal of understanding the
formation process for elliptical galaxies and the bulges of spiral
galaxies has been advanced by the quantification of the percentages
of counter-rotating cores, rotationally supported disks and the
measurement of spatially resolved stellar ages and abundances in
these systems.

As illustrated in the conference poster and also in
\citet{2004AN....325..100M} and \citet{2004A&A...415..889C}, higher
spatial resolution observations of galactic cores often brings to
light substructure that was invisible at coarser resolution. As with
most of the cases in this review, the greatly increased sky coverage
of a LGS system is needed to enable observations of a reasonable
sample.

\subsection{Nearby Active Galactic Nuclei}
\label{subsec-agn}

I do not propose to comment at any length on the use of OASIS+NAOMI
with a LGS to study Active Galactic Nuclei (AGN) either, as this was
also covered by several other speakers at the workshop. However, I
could not resist a nostalgic revisit to a paper from twenty years
ago (\citet{1985MNRAS.216..193M}) which I claim already showed both
the power of 3D spectroscopy for unravelling the complex dynamics of
gas in the inner regions of AGN, and also the difficulty in using
those measured dynamics to make general conclusions.

I would also claim that \citet{2004MNRAS.352.1180F}, who present
some recent OASIS data, faced the same problem. Indeed, a pessimist
might conclude that a large fraction of the observed gas dynamics in
AGN cores is chaotic, and, like the weather on the Earth, driven by
the (AGN equivalent of) flaps of butterfly wings in China.

In order not to leave this topic on a wholly negative note, it is
true that a similar conclusion might have been drawn about the
stellar dynamics discussed in section \ref{subsec-sauron}. As shown
by the papers referred to there, large enough samples of high
quality 3D data may allow more general and useful conclusions to be
drawn.

Some of the first AO assisted IFU observations were of `the usual
suspects' AGN - NGC 4151 and NGC 1068. These targets can be used as
their own guide stars and are hence well studied. I believe that
until a sample of size comparable to the SAURON galaxy survey has
been observed, at a similar level of detail, `general and useful
conclusions' will remain elusive. There is also a well known link
between AGN activity and star formation (and corresponding star
formation driven winds). These two phenomena are so closely
inter-related that it would seem foolish to me to study one without
studying the other. It has always been one of the great strengths
(and weaknesses) of 3D spectroscopy that, with a single observation,
one generally gets more information than can easily be dealt with on
all of the phenomena in the region observed. For this reason, this
approach may be the only one able to discover some underlying
physical principles hidden amongst the complex phenomenology of the
inner regions of AGN.

\section{Examples of Distant-Galaxy Science Enabled by a LGS}
\label{sec-distgal}

At first sight, one might suspect that using a 4m class telescope
with fine spatial sampling to observe extremely distant (and hence
faint) objects spectroscopically would be doomed to failure. In
general, this suspicion is in fact true, and most of the results
presented below are from 8m class telescopes. Two factors act to
open up a significant expanse of `discovery space' for OASIS+NAOMI
with a LGS. The first is that, in {\bf emission lines} many
extragalactic objects have regions of high surface brightness, and
second, that many of these regions are in fact small, or knotty, so
that any improvement in image quality over natural seeing can, to a
certain extent, compensate for a smaller telescope collecting area.

\subsection{Extreme Star Formation}
\label{subsec-scuba}

Observations in the sub-mm region of the spectrum have shown that
there is a significant population of dusty galaxies forming stars at
extra-ordinary rates ($\approx 1000\ M_{\odot}\ yr^{-1}$). Key
questions for these objects are: Are these massive galaxies with
extreme star-formation rates, or simply an extremely active phase in
more mundane systems? What drives the immense luminosity? What are
the dynamical masses? What are the metallicities of the various
components?

\citet{2005MNRAS.359..401S} demonstrate that a combination of
optical and Near-IR (NIR) IFU observations can begin to unravel the
answers to these questions. HST imaging has already also
demonstrated that the spatial scales of interest are significantly
less than an arcsecond.

More locally, \citet{2005ApJ...622..260S} show that optical IFU
observations of galaxies with spectroscopic evidence for a recent
cessation of rapid star formation may help us to determine the cause
of this truncation. Although probably a heterogeneous group,
galaxies with strong Balmer absorption features in their spectra
(from A stars), but weak emission lines indicating an absence of O
and B stars, likely include objects caught in transition between
star forming `spirals' and passive `ellipticals'. High spatial
resolution absorption line spectroscopy will no doubt be extremely
time consuming on a 4m class telescope, but is definitely not
impossible, and the spatial resolution of an LGS system should
deliver many of the benefits outlined in section \ref{subsec-sauron}
for the SAURON project.

\subsection{Gravitational Lensed High Redshift Galaxies}
\label{subsec-imbh}

One of the images used in the original science case for a LGS for
the WHT was that of the gravitationally multiply-lensed galaxy
behind the cluster Cl 0024+1654
(http://hubblesite.org/newscenter/newsdesk/archive/releases/1996/10/).
The gravitational lens increases the total flux collected from a
distant galaxy and spreads the unlensed image out over one or more
magnified images. Unfortunately, such fortuitous alignments of
galaxies with locations of strong lensing are rare, and, again,
asking for the additional requirement of a bright natural guide star
reduces the sample which can be studied with high spatial resolution
to near zero. The cluster Cl 0024+1654 is a case in point, with the
nearest available star 65 arcsec away and 16.5 magnitude in R
(making it a suitable tip-tilt reference for LGS AO, but of no use
for NGS AO).

\citet{2003ApJ...598..162S} demonstrated the power of this approach
on a z$\approx$1 galaxy, where they show that by using an IFU, and
correcting for the distortions in the image by the lensing process,
one can derive accurate rotation curves, and hence mass-estimates at
high redshifts. These measurements also allow the construction of
Tully-Fisher diagrams at times when the Universe was half its
present age. These in turn can be used to test models for galaxy
formation.

\subsection{Lyman $\alpha$ Emitters}
\label{subsec-lya}

Objects with redshifts between 3 and 7 have rest frame Lyman
$\alpha$ at observed wavelengths convenient for ground based AO
assisted IFU observations.

The SAURON IFU spectrograph on the WHT has been used on some lower
redshift (z$\approx$3) strong Lyman $\alpha$ emitters, deriving some
exciting conclusions about feedback from star formation at high
redshift \citep{2004MNRAS.351...63B, 2005Nat....436.227V}. Although
extremely long exposure times may be needed, LGS AO assisted IFU
observations of these, and other similar objects, could greatly
improve our understanding of the gas dynamics and the radiative
transfer of Lyman $\alpha$ photons in galaxies. As an example, in
the Lyman Break Galaxy (LBG) C15 discussed in
\citet{2004MNRAS.351...63B}, a clear velocity shear can be seen
across the line emitting region (even with seeing limited spatial
resolution). At the moment, the interpretation of that shear remains
ambiguous. Are we seeing an outflow tilted to our line of sight, or
are we seeing coherent rotation, holding out the promise of a total
mass estimate for this object?

\begin{figure*}
\centerline{\psfig{figure=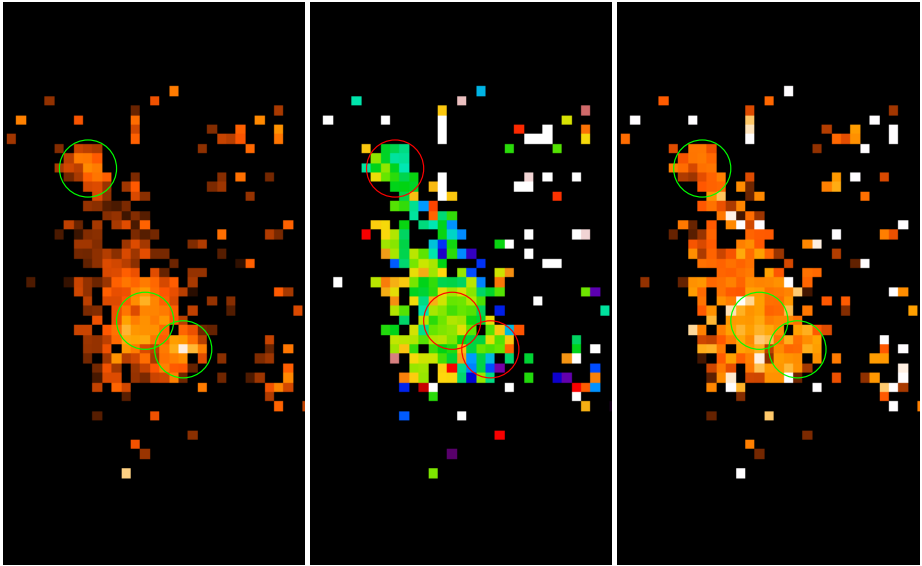,height=3in,angle=270}}
\caption{Single Gaussian fits to SAURON observations of the Steidel
`blob 1'. The region plotted is 41$\times$33 arcseconds in angular
size. Panel (a) shows the intensity of the fitted line (red--white:
0--$2\times10^{-17}$ ergs per cm$^{2}$ per sq.\ arcsec); Panel (b),
the central wavelength of the line (blue--red: 4970--4990\AA); Panel
(c), the width of the line (red--white: $\sigma = 0$--15\AA). The
plots allow us to quantify the velocity structure seen in the halo.
The circles show the regions of interest
\citep[see][]{2004MNRAS.351...63B}, with the top left circle
identifying the LBG C15 referred to in section \ref{subsec-lya}.}
\end{figure*}

In several knots of the large Lyman $\alpha$ emitter studied by
\citet{2005Nat....436.227V}, there is strong absorption which they
interpret as evidence for an outflowing wind covering the whole
source. Higher spectral and spatial resolution observations,
although no doubt extremely expensive in observing time, might be
able to confirm this interpretation.

\section{Conclusions}
\label{sec-conc}

I hope this review has demonstrated that despite its now `small'
collecting area, and the traditionally photon starved nature of IFU
observations, there is still a substantial range of interesting
astrophysical questions that can be addressed using the LGS with
NAOMI and OASIS on the WHT. The combined gains from focussing on
high surface brightness emission line objects and increasing that
surface brightness by image sharpening with AO will keep the WHT
scientifically competitive on targets with no natural guide star.

I would like to end with a comment on serendipity. This might seem
unlikely to be relevant for a small field of view, high spatial
resolution instrument like OASIS, but I would claim that, to a first
approximation, the chances of serendipitous discoveries are related
to the number of independent spatial and spectral samples, and the
uniqueness of the planned targets of the instrument. Some historical
examples that demonstrate this, despite having relatively small sky
coverage were published by \citet{1988ApJ...328L..29M,
1998ApJ...498L..93D}. A recent IFU example of serendipity
(admittedly from a `wide field' IFU) was published by
\citet{2005MNRAS.358L..11J}. OASIS can sample a substantial volume
of space, if one views its wavelength axis as a redshift range, and
will be one of the very few instruments able to deliver AO image
quality over most of the sky. In amongst all the careful discussions
of signal-to-noise and detection limits, it is good to also remember
how exciting astronomical discovery is, and how often astronomers
have been surprised by what they have found.


\end{document}